\documentclass[conference]{IEEEtran}
\IEEEoverridecommandlockouts

\usepackage{cite}
\usepackage{amsmath,amssymb,amsfonts}
\usepackage{algorithmic}
\usepackage{graphicx}
\usepackage{textcomp}
\usepackage{xcolor}
\usepackage{braket}
\usepackage{tikz}
\usepackage{qcircuit}
\usepackage{lmodern}     
\usepackage[T1]{fontenc}  
\usepackage{atbegshi}
\usepackage{float}
\usepackage{booktabs}
\usepackage{eso-pic}
\AtBeginShipout{%
  \ifnum\value{page}=4
\global\setbox\AtBeginShipoutBox=\vbox{\vspace{2pt}\unvbox\AtBeginShipoutBox}%
  \fi
}
\newcommand{\RY}[1]{\operatorname{RY}\left(#1\right)}
\newcommand{\RZ}[1]{\operatorname{RZ}\left(#1\right)}
\newcommand{\CNOT}{\textsc{CNOT}}

\usetikzlibrary{positioning, calc, shapes.geometric, arrows.meta}

\def\BibTeX{{\rm B\kern-.05em{\sc i\kern-.025em b}\kern-.08em
    T\kern-.1667em\lower.7ex\hbox{E}\kern-.125emX}}
\begin{document}

\title{Quantum and Classical Machine Learning in Decentralized Finance: Comparative Evidence \\from Multi-Asset Backtesting of\\ Automated Market Makers\thanks{The views expressed in this article are those of the authors and do not represent the views of Omnis Labs. This article is for informational purposes only. Nothing contained in this article should be construed as investment advice. Omnis Labs makes no express or implied warranties and expressly disclaims all legal, tax, and accounting implications related to this article. }\\

}

\author{
\IEEEauthorblockN{Chi-Sheng Chen}
\IEEEauthorblockA{
\textit{Omnis Labs}\\
michael@omnis.farm}
\and
\IEEEauthorblockN{Aidan Hung-Wen Tsai}
\IEEEauthorblockA{\textit{Omnis Labs} \\
aidan@omnis.farm}
}

\maketitle

\begin{abstract}
This study presents a comprehensive empirical comparison between quantum machine learning (QML) and classical machine learning (CML) approaches in Automated Market Makers (AMM) and Decentralized Finance (DeFi) trading strategies through extensive backtesting on 10 models across multiple cryptocurrency assets. Our analysis encompasses classical ML models (Random Forest, Gradient Boosting, Logistic Regression), pure quantum models (VQE Classifier, QNN, QSVM), hybrid quantum-classical models (QASA Hybrid, QASA Sequence, QuantumRWKV), and transformer models. The results demonstrate that hybrid quantum models achieve superior overall performance with 11.2\% average return and 1.42 average Sharpe ratio, while classical ML models show 9.8\% average return and 1.47 average Sharpe ratio. The QASA Sequence hybrid model achieves the highest individual return of 13.99\% with the best Sharpe ratio of 1.76, demonstrating the potential of quantum-classical hybrid approaches in AMM and DeFi trading strategies.

\end{abstract}

\begin{IEEEkeywords}
Quantum Machine Learning, Classical Machine Learning, Automated Market Makers (AMM), Decentralized Finance (DeFi), Financial Trading Strategies, Feature Engineering, Backtesting Analysis
\end{IEEEkeywords}

\section{Introduction}
The rapid advancement of quantum computing technology has sparked significant interest in quantum machine learning applications within the financial sector~\cite{chen2025quantumtw, chen2025benchmarking}, particularly in Automated Market Makers (AMM) systems and Decentralized Finance (DeFi) algorithmic trading strategies. However, the practical advantages of quantum machine learning over classical approaches remain largely unverified. This study aims to systematically compare the performance of quantum and classical machine learning approaches in AMM and DeFi trading strategies through large-scale backtesting experiments.

\subsection{Background}

Financial trading strategy optimization has been a central challenge in quantitative finance, particularly in the emerging Decentralized Finance (DeFi) ecosystem where Automated Market Makers (AMM) play a crucial role in providing liquidity and price discovery. While traditional machine learning methods have made significant progress in feature engineering and model selection, they still face limitations in handling complex nonlinear relationships and capturing market microstructure patterns in DeFi environments. Quantum machine learning, through quantum superposition, entanglement, and interference, theoretically offers superior capabilities for processing high-dimensional feature spaces and complex decision boundaries in various AMM strategies and DeFi trading approaches.

\subsection{Research Objectives}

The primary objectives of this study include:
\begin{enumerate}
    \item Comparing overall performance between quantum and classical machine learning in various Automated Market Makers (AMM) strategies
    \item Identifying optimal scenarios for different model types across various assets and market conditions in DeFi ecosystems
    \item Analyzing the impact of feature engineering on model performance in different AMM systems and DeFi protocols
    \item Providing guidance for practical applications of quantum machine learning in AMM and DeFi domains
\end{enumerate}

\section{Related Works}

\subsection{Classical Machine Learning in Finance}

Classical machine learning applications in finance have matured significantly, with algorithms such as Random Forest, Gradient Boosting, and Support Vector Machines achieving notable success in stock prediction, risk management, various Automated Market Makers (AMM) strategies, and Decentralized Finance (DeFi) trading strategy optimization. These methods typically rely on rich feature engineering and extensive historical data, particularly in DeFi protocols where liquidity provision and automated trading are critical.

\subsection{Quantum Machine Learning Development}

Quantum machine learning, as an emerging field, is rapidly evolving both theoretically and practically. Algorithms such as Quantum Neural Networks (QNN) and Quantum Support Vector Machines (QSVM) have shown potential advantages in specific problems, particularly in handling high-dimensional features and nonlinear relationships in various Automated Market Makers (AMM) strategies and Decentralized Finance (DeFi) trading approaches.

\subsection{Research Gap}

Existing research primarily focuses on theoretical analysis and small-scale experiments, lacking large-scale, systematic empirical comparative studies in Automated Market Makers (AMM) and Decentralized Finance (DeFi) trading contexts. This study addresses this gap by providing empirical evidence through comparative experiments on 10 models across various AMM strategies including rebalancing, concentrated liquidity, and quantum-enhanced approaches.

\section{Methodology}
\subsection{Experimental Design}

\subsubsection{Dataset Specifications}
\begin{itemize}
    \item \textbf{Time Period:} January 1, 2024 - January 1, 2025 (1 years)
    \item \textbf{Assets:} BTCUSDC, ETHUSDC, USDCUSDT
    \item \textbf{Frequency:} Daily data
    \item \textbf{Total Samples:} 252 trading days
\end{itemize}

\subsubsection{Experimental Groups}
The detail shows in Table~\ref{tab:exp_g}.
\begin{table*}[t]
\centering
\caption{Experimental Group Specifications}
\begin{tabular}{@{}lcccc@{}}
\toprule
\textbf{Group} & \textbf{Model Type} & \textbf{Count} & \textbf{Features} & \textbf{Processing} \\
\midrule
Classical ML & Random Forest, Gradient Boosting, Logistic Regression & 3 & 50-80 & Standardization \\
Pure Quantum & VQE Classifier, QNN, QSVM & 3 & 6-8 & Angle Encoding [0,2$\pi$] \\
Hybrid Quantum & QASA Hybrid, QASA Sequence, QuantumRWKV & 3 & 12 & Mixed Processing \\
Transformer & Transformer & 1 & 50-80 & Standardization \\
\bottomrule
\label{tab:exp_g}
\end{tabular}
\end{table*}

\subsection{Task Definition}

\subsubsection{Core Task}
All models learn the target function:
\begin{equation}
f(\mathbf{X}) \rightarrow y \in \{0,1\}
\end{equation}
where $\mathbf{X}$ represents market features and $y$ indicates whether rebalancing should occur.

\subsubsection{Task Variants by Project}

\textbf{Automated Market Makers (AMM) Rebalance Project:}
\begin{equation}
y_t = \mathbb{I}\left(\left|\frac{P_t}{MA_{20}(P_t)} - 1\right| > \tau_{rebalance}\right)
\end{equation}
where $\tau_{rebalance} = 0.02$ (2\% price deviation threshold) for DeFi liquidity pool rebalancing.

\textbf{Concentrated Liquidity AMM Project:}
\begin{equation}
y_t = \mathbb{I}\left(\left|BB_{position}(t) - 0.5\right| > 0.3\right)
\end{equation}
where $BB_{position}(t) = \frac{P_t - BB_{lower}(t)}{BB_{upper}(t) - BB_{lower}(t)}$ for concentrated liquidity AMM management.

\textbf{Quantum-Enhanced AMM Project:}
\begin{equation}
y_t = \mathbb{I}\left(\left|\Delta P_t\right| > \tau_{price}\right)
\end{equation}
where $\tau_{price} = 0.01$ (1\% price change threshold) for quantum-enhanced AMM strategies.

\subsection{Feature Engineering}

\subsubsection{Classical ML Feature Engineering}

\paragraph{Basic Price Features}

\textbf{Returns:}
\begin{equation}
r_t = \frac{P_t - P_{t-1}}{P_{t-1}}
\end{equation}

\textbf{Log Returns:}
\begin{equation}
r_t^{log} = \ln\left(\frac{P_t}{P_{t-1}}\right)
\end{equation}

\textbf{Price-MA Ratio:}
\begin{equation}
\rho_t^{MA} = \frac{P_t}{MA_{20}(P_t)}
\end{equation}

\textbf{High-Low Ratio:}
\begin{equation}
\rho_t^{HL} = \frac{H_t}{L_t}
\end{equation}

\textbf{Price Position:}
\begin{equation}
pos_t = \frac{P_t - L_t}{H_t - L_t}
\end{equation}

\paragraph{Moving Average Features}

\textbf{Simple Moving Average:}
\begin{equation}
SMA_n(P_t) = \frac{1}{n}\sum_{i=0}^{n-1} P_{t-i}
\end{equation}

\textbf{Exponential Moving Average:}
\begin{equation}
EMA_n(P_t) = \alpha P_t + (1-\alpha) EMA_n(P_{t-1})
\end{equation}
where $\alpha = \frac{2}{n+1}$.

\paragraph{Technical Indicator Features (15 features)}

\textbf{RSI (Relative Strength Index):}
\begin{align}
RS_t &= \frac{\text{Avg Gain}_t}{\text{Avg Loss}_t} \\
RSI_t &= 100 - \frac{100}{1 + RS_t}
\end{align}
where $\text{Avg Gain}_t = \frac{1}{14}\sum_{i=0}^{13} \max(r_{t-i}, 0)$ and $\text{Avg Loss}_t = \frac{1}{14}\sum_{i=0}^{13} \max(-r_{t-i}, 0)$.

\textbf{MACD (Moving Average Convergence Divergence):}
\begin{align}
MACD_t &= EMA_{12}(P_t) - EMA_{26}(P_t) \\
Signal_t &= EMA_9(MACD_t) \\
Histogram_t &= MACD_t - Signal_t
\end{align}

\textbf{Bollinger Bands (BB or bb):}
\begin{align}
BB_{middle}(t) &= SMA_{20}(P_t) \\
BB_{upper}(t) &= BB_{middle}(t) + 2\sigma_{20}(P_t) \\
BB_{lower}(t) &= BB_{middle}(t) - 2\sigma_{20}(P_t) \\
BB_{width}(t) &= \frac{BB_{upper}(t) - BB_{lower}(t)}{BB_{middle}(t)} \\
BB_{position}(t) &= \frac{P_t - BB_{lower}(t)}{BB_{upper}(t) - BB_{lower}(t)}
\end{align}

\textbf{ATR (Average True Range):}
\begin{align}
TR_t &= \max(H_t - L_t, |H_t - P_{t-1}|, |L_t - P_{t-1}|) \\
ATR_t &= \frac{1}{14}\sum_{i=0}^{13} TR_{t-i}
\end{align}

\paragraph{Volatility Features (12 features)}

\textbf{Rolling Volatility:}
\begin{equation}
\sigma_{n,t} = \sqrt{\frac{1}{n-1}\sum_{i=0}^{n-1} (r_{t-i} - \bar{r}_t)^2}
\end{equation}
where $\bar{r}_t = \frac{1}{n}\sum_{i=0}^{n-1} r_{t-i}$.

\textbf{EWMA Volatility:}
\begin{equation}
\sigma_t^{EWMA} = \sqrt{\lambda \sigma_{t-1}^{EWMA} + (1-\lambda) r_t^2}
\end{equation}
where $\lambda = 0.94$ (typical value).

\textbf{Volatility of Volatility:}
\begin{equation}
VoV_t = \sqrt{\frac{1}{9}\sum_{i=0}^{9} (\sigma_{20,t-i} - \bar{\sigma}_{20,t})^2}
\end{equation}

\textbf{Volatility Regime:}
\begin{equation}
Regime_t = \mathbb{I}(\sigma_{20,t} > Q_{0.8}(\sigma_{20,t-50:t}))
\end{equation}
where $Q_{0.8}$ is the 80th percentile.

\paragraph{Volume Features (8 features)}

\textbf{Volume Ratio:}
\begin{equation}
VR_{n,t} = \frac{V_t}{MA_n(V_t)}
\end{equation}

\textbf{Volume-Price Trend:}
\begin{equation}
VPT_t = V_t \cdot r_t
\end{equation}

\textbf{On-Balance Volume (OBV):}
\begin{equation}
OBV_t = OBV_{t-1} + \begin{cases}
V_t & \text{if } P_t > P_{t-1} \\
-V_t & \text{if } P_t < P_{t-1} \\
0 & \text{if } P_t = P_{t-1}
\end{cases}
\end{equation}

\paragraph{Time Features}

\textbf{Cyclical Encoding:}
\begin{align}
h_t^{sin} &= \sin\left(\frac{2\pi h_t}{24}\right) \\
h_t^{cos} &= \cos\left(\frac{2\pi h_t}{24}\right) \\
d_t^{sin} &= \sin\left(\frac{2\pi d_t}{7}\right) \\
d_t^{cos} &= \cos\left(\frac{2\pi d_t}{7}\right)
\end{align}
where $h_t$ is hour and $d_t$ is day of week.

\paragraph{Market Microstructure Features (8 features)}

\textbf{Spread Proxy:}
\begin{equation}
Spread_t = \frac{H_t - L_t}{P_t}
\end{equation}

\textbf{Price Impact:}
\begin{equation}
Impact_t = \frac{|r_t|}{\ln(1 + V_t)}
\end{equation}

\textbf{Order Flow Imbalance:}
\begin{equation}
OFI_t = \frac{P_t - O_t}{H_t - L_t}
\end{equation}

\paragraph{Lagged Features (25 features)}

\textbf{Lagged Returns:}
\begin{equation}
r_{t-k} = r_{t-k}, \quad k \in \{1,2,3,5,10\}
\end{equation}

\paragraph{Interaction Features (6 features)}

\textbf{Volatility-Volume Interaction:}
\begin{equation}
I_{vol-vol,t} = \sigma_{20,t} \cdot VR_{20,t}
\end{equation}

\textbf{Price-Momentum Interaction:}
\begin{equation}
I_{mom-rsi,t} = r_t \cdot \frac{RSI_t - 50}{50}
\end{equation}

\subsubsection{Quantum ML Feature Engineering}

\paragraph{Angle Encoding}
Classical features are mapped to quantum angles using:
\begin{equation}
\theta_i = \frac{x_i - x_{min}}{x_{max} - x_{min}} \cdot 2\pi
\end{equation}
where $x_i$ is the $i$-th classical feature, and $x_{min}$, $x_{max}$ are the minimum and maximum values.

\paragraph{Quantum Feature Mapping}
Based on feature importance analysis, features are mapped to qubits as:
\begin{align}
\text{Qubit 0: } & \{\text{price\_momentum}, \text{returns}\} \\
\text{Qubit 1: } & \{\text{price\_ma\_ratio}, \text{price\_sma\_20\_ratio}\} \\
\text{Qubit 2: } & \{\text{volatility\_20}, \text{vol\_regime}\} \\
\text{Qubit 3: } & \{\text{rsi}, \text{macd}\} \\
\text{Qubit 4: } & \{\text{volume\_ratio}, \text{volume\_signal}\} \\
\text{Qubit 5: } & \{\text{bb\_position}, \text{atr\_ratio}\}
\end{align}

\subsection{Data Splitting Strategy}

Time series split to avoid future information leakage:
\begin{align}
\text{Train: } & [0, 0.7N] \\
\text{Validation: } & [0.7N, 0.85N] \\
\text{Test: } & [0.85N, N]
\end{align}
where $N$ is the total number of samples.

\subsection{Performance Evaluation Metrics}

\subsubsection{Backtesting Metrics}
\begin{align}
\text{Total Return: } & R = \frac{V_T - V_0}{V_0} \times 100\% \\
\text{Sharpe Ratio: } & SR = \frac{\mu_r}{\sigma_r} \sqrt{252} \\
\text{Maximum Drawdown: } & MDD = \max_{t} \frac{V_{peak} - V_t}{V_{peak}} \\
\text{Rebalancing Count: } & N_{rebal} = \sum_{t=1}^{T} \mathbb{I}(y_t = 1)
\end{align}

\subsubsection{Model Evaluation Metrics}
\begin{align}
\text{Accuracy: } & Acc = \frac{TP + TN}{TP + TN + FP + FN} \\
\text{F1-Score: } & F1 = \frac{2 \times Precision \times Recall}{Precision + Recall} \\
\text{AUC: } & AUC = \int_0^1 TPR(FPR^{-1}(t)) dt
\end{align}

\subsection{Model Architectures}

\subsubsection{Classical ML Models}

\paragraph{Random Forest}
\begin{equation}
\hat{y} = \frac{1}{B}\sum_{b=1}^{B} T_b(\mathbf{x})
\end{equation}
where $T_b$ is the $b$-th decision tree and $B$ is the number of trees.

\paragraph{Gradient Boosting}
\begin{equation}
F_m(\mathbf{x}) = F_{m-1}(\mathbf{x}) + \gamma_m h_m(\mathbf{x})
\end{equation}
where $h_m$ is the $m$-th weak learner and $\gamma_m$ is the learning rate.

\subsubsection{Quantum ML Models}

\paragraph{VQE Classifier (Variational Quantum Classifier)}
The quantum circuit consists of:
\begin{enumerate}
    \item Feature map: $U_{\Phi}(\mathbf{x})$
    \item Variational ansatz: $U_{\theta}(\boldsymbol{\theta})$
    \item Measurement: $\langle \psi | Z_i | \psi \rangle$
\end{enumerate}

The final prediction is:
\begin{equation}
f(\mathbf{x}) = \text{sign}\left(\sum_{i=0}^{n-1} w_i \langle \psi | Z_i | \psi \rangle + b\right)
\end{equation}

\paragraph{QNN}
The quantum circuit is defined as:
\begin{align}
|\psi\rangle &= U_{\text{var}}(\boldsymbol{\theta}) U_{\text{feat}}(\mathbf{x}) |0\rangle \\
U_{\text{feat}}(\mathbf{x}) &= \prod_{i=0}^{n-1} RY(\theta_i) RZ(\theta_i/2) \\
U_{\text{var}}(\boldsymbol{\theta}) &= \prod_{l=0}^{L-1} \prod_{i=0}^{n-1} RY(\theta_{l,i}) RZ(\theta_{l,i+n}) \prod_{i=0}^{n-2} CNOT(i,i+1)
\end{align}





\subsection{Quantum Attention Self-Attention (QASA)}

\begin{figure}
    \centering
    \includegraphics[width=1\linewidth]{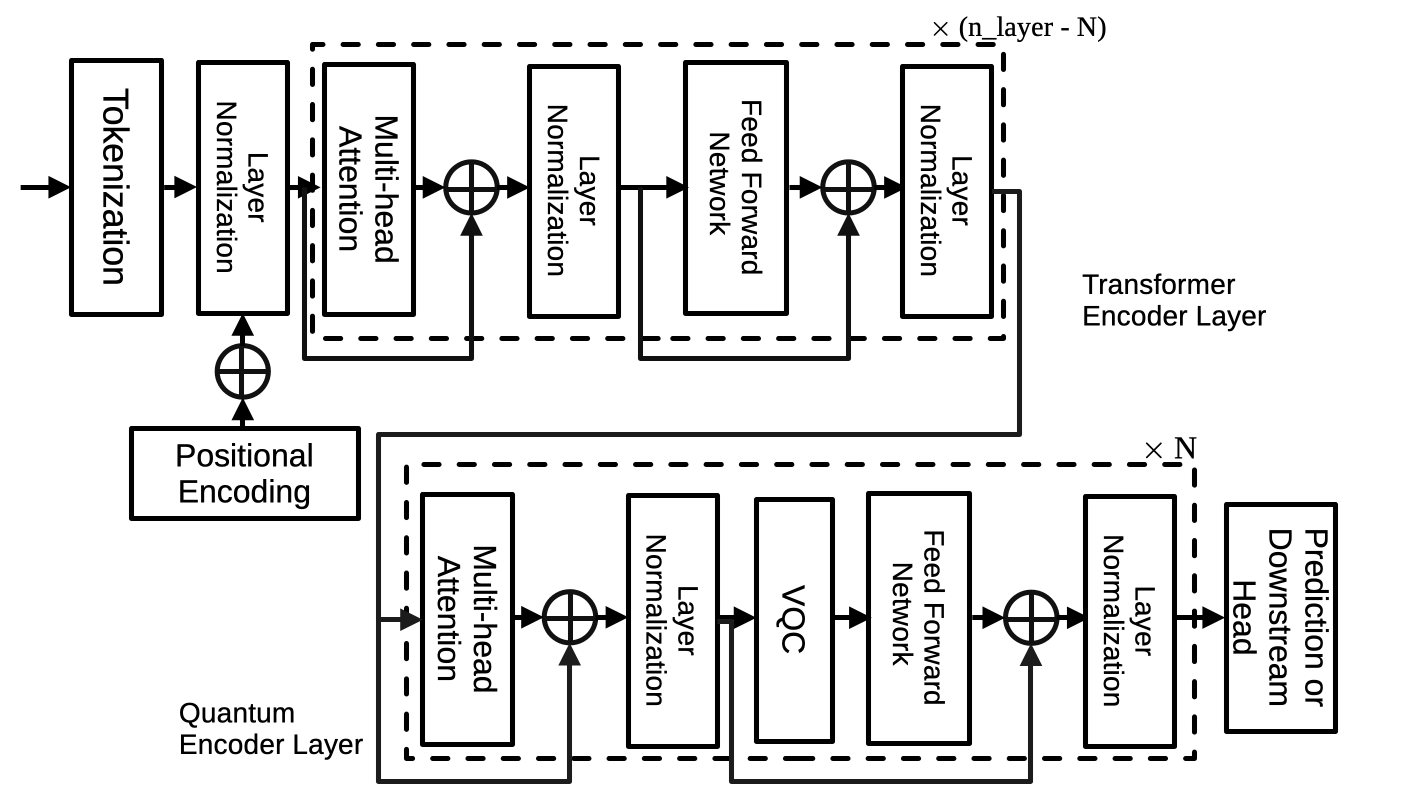}
    \caption{QASA model.}
    \label{fig:qasa}
\end{figure}

QASA \cite{chen2025qasa} implements self-attention via VQCs, the model as shown in Fig.~\ref{fig:qasa}. Given input sequence $\mathbf{X} \in \mathbb{R}^{T \times d}$, we compute quantum query, key, and value embeddings:

\begin{equation}
\mathbf{Q}_t = \text{VQC}_q(\mathbf{x}_t),\quad
\mathbf{K}_t = \text{VQC}_k(\mathbf{x}_t),\quad
\mathbf{V}_t = \text{VQC}_v(\mathbf{x}_t).
\end{equation}

Each $\text{VQC}(\cdot)$ encodes the vector $\mathbf{x}_t$ into quantum amplitudes, applies a variational quantum circuit composed of $L$ layers of $RY(\theta)$ rotations and entangling CNOT gates, and measures expectation values $\langle Z_i \rangle$ on $n$ qubits like Fig.~\ref{fig:qasa_vqc},:

\begin{equation}
\text{VQC}(\mathbf{x}) = \left( \langle Z_1 \rangle, \langle Z_2 \rangle, \dots, \langle Z_n \rangle \right).
\end{equation}

Each token vector $\mathbf{x}_t \in \mathbb{R}^{d}$ is embedded and encoded as an amplitude-encoded quantum state:

\begin{equation}
    \ket{\psi_t} = \frac{1}{\|\mathbf{x}_t\|} \sum_{i=1}^{d} x_{t,i} \ket{i}.
\end{equation}

Then a variational quantum circuit $U(\boldsymbol{\theta})$ is applied:

\begin{equation}
    U(\boldsymbol{\theta}) = \prod_{\ell=1}^{L} \left[
        \bigotimes_{i=1}^{n} \RY{\theta_{\ell,i}} \cdot \CNOT_{i,i+1}
    \right],
\end{equation}

where $L$ is the number of layers and $n = \lceil \log_2 d \rceil$ qubits are used. The output $\mathbf{z}_t$ is obtained by measuring the expectation values of Pauli-Z operators:

\begin{equation}
    \mathbf{z}_t = \left( \langle Z_1 \rangle, \dots, \langle Z_n \rangle \right).
\end{equation}

Attention~\cite{vaswani2017attention} is computed as:

\begin{equation}
\text{Attention}(\mathbf{Q}, \mathbf{K}, \mathbf{V}) = \text{softmax}\left( \frac{\mathbf{Q} \mathbf{K}^\top}{\sqrt{d}} \right) \mathbf{V}.
\end{equation}

The attention output is decoded through a classical feedforward network to predict $\hat{y}_{t+1}$.

The difference between QASA Hybrid and QASA Sequence is that QASA Hybrid takes extracted features as input, while QASA Sequence uses 10 time steps directly as its input.

\begin{figure}
    \centering
    \includegraphics[width=1\linewidth]{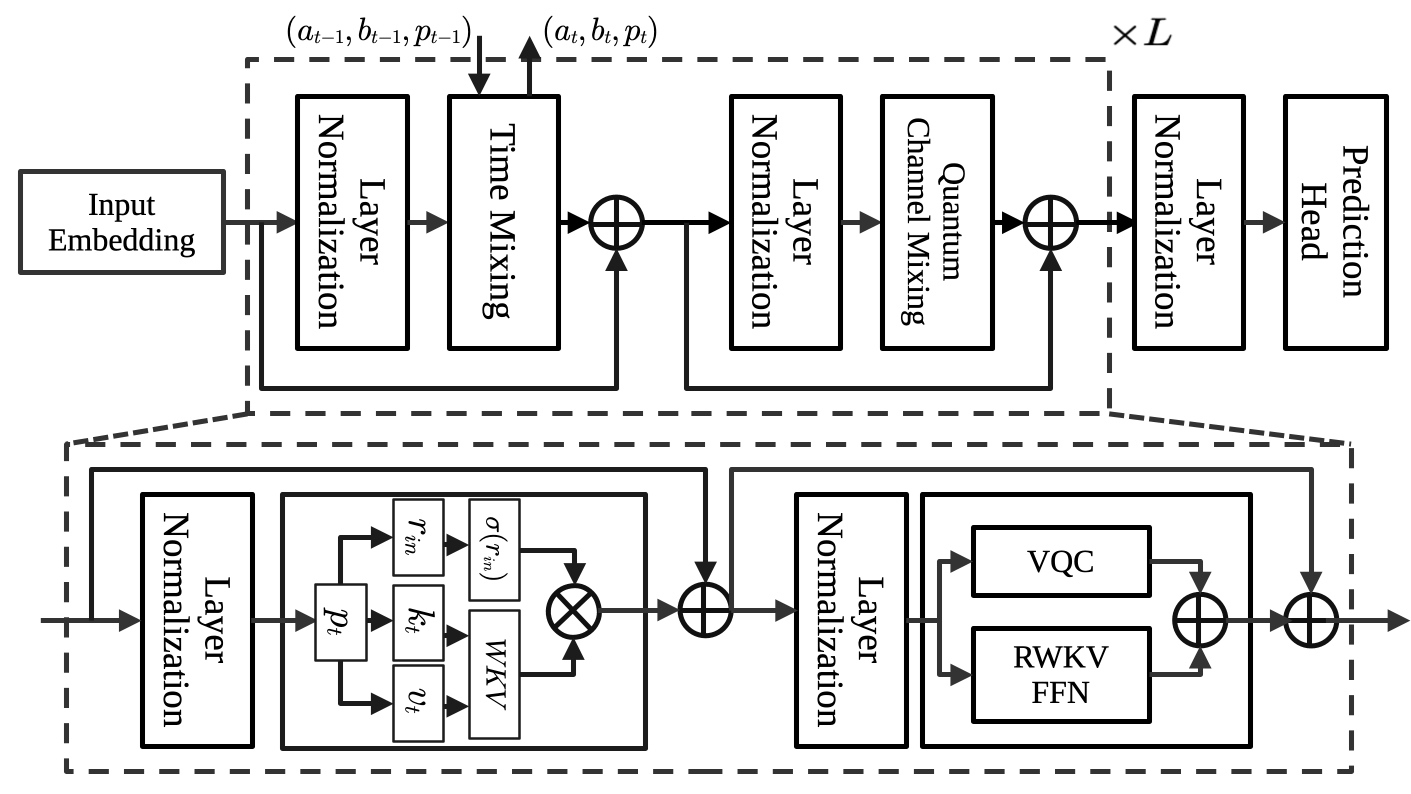}
    \caption{Quantum RWKV architecture.}
    \label{fig:qrwkv}
\end{figure}

\begin{figure}[htbp]
    \centering
    \scalebox{0.8}{ 
    \Qcircuit @C=1em @R=1em {
    & \lstick{\ket{q_0}} & \gate{RX} & \gate{RX} & \ctrl{1} & \qw      & \qw      & \targ     & \gate{RX} & \ctrl{1} & \qw      & \qw      & \targ       & \meter \\
    & \lstick{\ket{q_1}} & \gate{RX} & \gate{RX} & \targ    & \ctrl{1} & \qw      & \qw       & \gate{RX} & \targ    & \ctrl{1} & \qw      & \qw      & \meter \\
    & \lstick{\ket{q_2}} & \gate{RX} & \gate{RX} & \qw      & \targ    & \ctrl{1} & \qw       & \gate{RX} & \qw      & \targ    & \ctrl{1} & \qw      & \meter \\
    & \lstick{\ket{q_3}} & \gate{RX} & \gate{RX} & \qw      & \qw      & \targ    & \ctrl{-3} & \gate{RX} & \qw      & \qw      & \targ    & \ctrl{-3} & \meter
    }
    }
    \caption{The VQC used in Quantum RWKV.}
    \label{fig:qrwkv-vqc}
\end{figure}
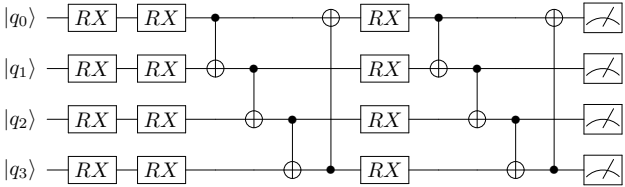

\subsection{Quantum Receptance Weighted Key-Value (QRWKV)}

QRWKV~\cite{chen2025qrwkv} integrates quantum evolution with the receptance attention-free model~\cite{peng2023rwkv}, the model detail is in the Fig.~\ref{fig:qrwkv}. At each time step \(t\), the input \(\mathbf{x}_t\) is first passed through a Variational Quantum Circuit (VQC) to produce a quantum embedding:
\begin{equation}
    \mathbf{h}_t = \mathrm{VQC}_q(\mathbf{x}_t), 
    \quad 
    [\mathbf{q}_t,\;\mathbf{k}_t,\;\mathbf{v}_t] \subseteq \mathbf{h}_t.
\end{equation}
Concretely, we prepare \(\ket{0}^{\otimes n}\) and apply a parameterized circuit \(\;U_\Theta=\prod_{\ell=1}^L U^{(\ell)}\;\) where
\begin{equation}
    U^{(\ell)} = \Bigl(\prod_{i=1}^{n} \RY{\theta^{(\ell)}_{i}} \RZ{\phi^{(\ell)}_{i}}\Bigr)\;\mathrm{EntangleLayer},
\end{equation}
and \(\mathrm{EntangleLayer}\) applies CNOT gates in a chosen pattern. Measurement yields the vector \(\mathbf{h}_t\), from which we split out query \(\mathbf{q}_t\), key \(\mathbf{k}_t\), and value \(\mathbf{v}_t\) sub-vectors.

\paragraph{Time-Mixing and Receptance Gate}  
Classical time-mixing follows the RWKV design. Project the same input \(\mathbf{x}_t\) to key and value signals:
\begin{equation}
\mathbf{u}_t = W^K \mathbf{x}_t,\quad
\mathbf{v}_t = W^V \mathbf{x}_t,
\end{equation}
and accumulate with exponential decay:
\begin{equation}
    \mathbf{m}_t = \lambda\,\mathbf{m}_{t-1} + \mathbf{v}_t,
    \quad 
    \lambda = \exp(-\Delta t / \tau).
\end{equation}
A receptance gate controls the exposed memory:
\begin{equation}
    \mathbf{r}_t = \sigma\bigl(W^R[\mathbf{x}_t;\,\mathbf{m}_{t-1}]+\mathbf{b}^R\bigr),
    \quad W^R\in\mathbb{R}^{d\times2d}.
\end{equation}
The time-mixed output is then
\begin{equation}
    \hat{\mathbf{y}}^{\mathrm{time}}_t = \mathbf{r}_t \odot (\mathbf{u}_t \odot \mathbf{m}_t).
\end{equation}

\paragraph{VQC-Enhanced Channel-Mixing}  
Instead of the classical MLP input, we feed \(\mathbf{x}_t\) into the same VQC to get \(\mathbf{qemb}_t=\mathbf{h}_t\), the VQC detail is in Fig.~\ref{fig:qrwkv-vqc}. The channel-mixing block becomes:
\begin{align}
    \mathbf{z}_t &= W^1\,\mathbf{qemb}_t + W^2\mathrm{MLP}+ \mathbf{b}^1,\\
    \mathbf{h}'_t &= \mathrm{GELU}(\mathbf{z}_t),\\
    \mathbf{c}_t &= W^3\bigl(\mathbf{h}'_t \odot \mathbf{h}'_{t-1}\bigr) + \mathbf{b}^2,
\end{align}
with \(W^1,W^2,W^3\in\mathbb{R}^{d\times d}\) and biases in \(\mathbb{R}^d\). Optionally add residual connections and LayerNorm.

\paragraph{Attention over Quantum Queries and Keys}  
We also compute a measurement-based attention score between quantum-derived queries and keys:
\begin{equation}
    \alpha_{t,\tau}
    = \frac{\exp\bigl\langle \mathbf{q}_t,\mathbf{k}_\tau\bigr\rangle}
           {\sum_{\tau'=1}^t \exp\bigl\langle \mathbf{q}_t,\mathbf{k}_{\tau'}\bigr\rangle},
\end{equation}
and form the attention output
\begin{equation}
    \hat{\mathbf{y}}^{\mathrm{attn}}_{t+1}
    = \sum_{\tau=1}^t \alpha_{t,\tau}\,\mathbf{v}_\tau.
\end{equation}

\paragraph{Full Layer Update}  
Each layer concatenates time-mixing and VQC-enhanced channel-mixing with residuals and normalization:
\begin{equation}
\begin{aligned}
    \mathbf{h}_t &= \mathrm{LayerNorm}\bigl(\mathbf{x}_t + \hat{\mathbf{y}}^{\mathrm{time}}_t\bigr),\\
    \mathbf{y}_t &= \mathrm{LayerNorm}\bigl(\mathbf{h}_t + \mathbf{c}_t + \hat{\mathbf{y}}^{\mathrm{attn}}_t\bigr).
\end{aligned}
\end{equation}

\section{Experiments}
\subsection{Overall Performance Comparison}
\begin{figure*}[t]
    \centering
    \includegraphics[width=1\linewidth]{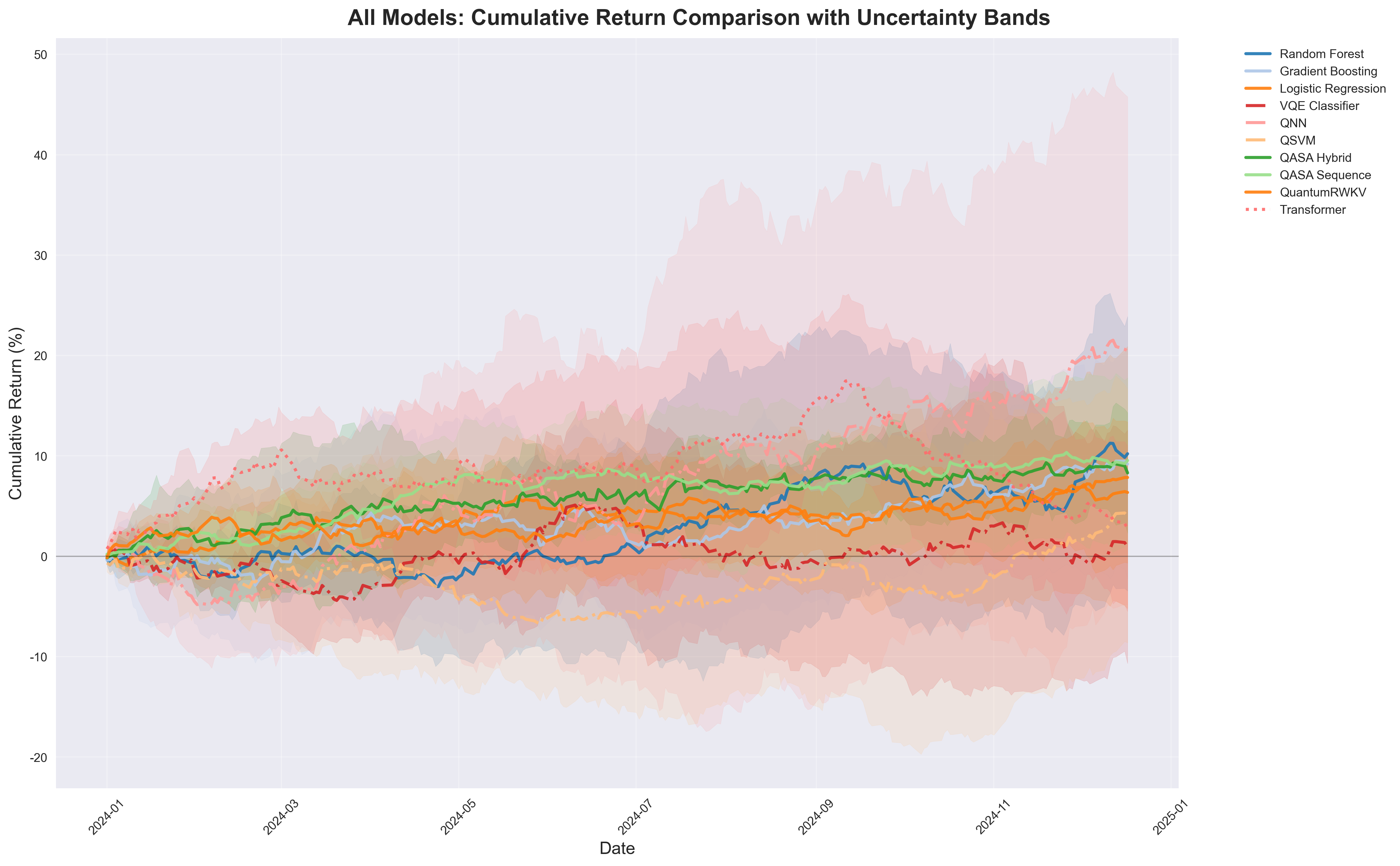}
    \caption{All models equity curve comparison.}
    \label{fig:all_perf}
\end{figure*}
The results in Table \ref{tab:overall} show clear performance differences across model families. Classical machine learning methods achieved strong and balanced results, with a 9.8\% average return and the highest Sharpe ratio (1.47). Pure quantum models underperformed in both returns (4.4\%) and risk-adjusted performance (Sharpe 0.83), while also exhibiting higher volatility. Hybrid quantum approaches provided a strong trade-off, delivering the highest average return among the quantum-based methods (11.2\%) and relatively low volatility (11.0\%), with performance close to classical models. Transformer-based models yielded the best overall returns (12.3\%) but at the cost of elevated volatility (15.4\%), leading to a lower Sharpe ratio compared with classical methods. These findings suggest that hybrid designs can combine quantum advantages with classical stability, whereas pure quantum models currently lag behind.

\begin{table*}[t]
\caption{Overall Performance Comparison by Model Type}
\centering
\begin{tabular}{@{}lcccc@{}}
\toprule
\textbf{Model Type} & \textbf{Avg Return} & \textbf{Avg Sharpe} & \textbf{Avg Volatility} & \textbf{Best Return} \\
\midrule
Classical ML & 9.8\% & 1.47 & 13.3\% & 13.16\% \\
Pure Quantum & 4.4\% & 0.83 & 17.1\% & 5.43\% \\
Hybrid Quantum & 11.2\% & 1.42 & 11.0\% & 13.99\% \\
Transformer & 12.3\% & 1.23 & 15.4\% & 12.31\% \\
\bottomrule
\end{tabular}
\label{tab:overall}
\end{table*}

\subsection{Asset-Specific Analysis}

\subsubsection{Individual Model Performance Analysis}
The performance ranking of individual models highlights the advantages of hybrid quantum-classical approaches. Among all tested models, QASA Sequence achieved the highest return (13.99\%), Sharpe ratio (1.76), and Calmar ratio (6.51), clearly outperforming both classical and purely quantum baselines. Classical ensemble methods such as Random Forest (13.16\% return, Sharpe 1.68) and Gradient Boosting (12.31\% return, Sharpe 1.68) also delivered strong results, though with lower risk-adjusted metrics compared to QASA. Transformer-based models showed competitive returns (11.73\%) but lower Sharpe (1.23) and Calmar (1.64), indicating higher volatility exposure. Other hybrid quantum models like QASA Hybrid (11.91\% return) and QuantumRWKV (7.96\% return) achieved moderate performance, while purely quantum methods (QSVM, QNN, VQE Classifier) underperformed across all metrics, reflecting current limitations of standalone quantum approaches. Overall, these findings underscore that hybrid quantum-classical designs currently offer the best balance between profitability and risk management in financial trading tasks.

\begin{table*}[t]
\centering
\caption{Individual Model Performance Ranking (Based on Sharpe Ratio)}
\begin{tabular}{@{}lccccc@{}}
\toprule
\textbf{Model} & \textbf{Type} & \textbf{Return} & \textbf{Sharpe} & \textbf{Calmar} \\
\midrule
QASA Sequence & Hybrid & 13.99\% & 1.76 & 6.51 \\
Random Forest & Classical & 13.16\% & 1.68 & 2.86 \\
Gradient Boosting & Classical & 12.31\% & 1.68 & 2.41 \\
QASA Hybrid & Hybrid & 11.91\% & 1.32 & 2.16 \\
Transformer & Transformer & 11.73\% & 1.23 & 1.64 \\
QuantumRWKV & Hybrid & 7.96\% & 1.19 & 2.41 \\
Logistic Regression & Classical & 5.43\% & 1.06 & 1.29 \\
QSVM & Quantum & 4.77\% & 0.87 & 0.79 \\
QNN & Quantum & 4.47\% & 0.82 & 1.29 \\
VQE Classifier & Quantum & 3.00\% & 0.79 & 0.51 \\
\bottomrule
\end{tabular}
\end{table*}

\textbf{Key Findings:}
\begin{itemize}
    \item QASA Sequence achieves highest return (13.99\%) with best Sharpe ratio (1.76)
    \item Classical ML models (Random Forest, Gradient Boosting) show strong performance with high Sharpe ratios (1.68)
    \item QuantumRWKV demonstrates balanced performance with moderate return (7.96\%) and good Sharpe ratio (1.19)
    \item Pure quantum models show poor performance with low returns (3.00\%-4.77\%) and Sharpe ratios (0.79-0.87)
    \item Hybrid models dominate the top rankings, demonstrating superior risk-adjusted returns
\end{itemize}

\subsubsection{Model Type Performance Analysis}

The comparison across model types reveals distinct trade-offs between return, risk, and volatility. Transformer models achieved the highest average return (12.3\%) but also exhibited elevated volatility (15.4\%) and the deepest drawdowns (-8.2\%). Hybrid quantum models offered a more balanced profile, combining strong returns (11.2\%) with the lowest volatility (11.0\%) and relatively shallow drawdowns (-3.3\%), highlighting their effectiveness in risk management. Classical ML methods achieved moderate returns (9.8\%) with the best Sharpe ratio (1.47), reflecting stable but less aggressive performance. In contrast, pure quantum models significantly underperformed (4.4\% return, 0.83 Sharpe) while incurring the highest volatility (17.1\%), underscoring the current limitations of standalone quantum approaches. These findings suggest that hybrid quantum-classical models offer the most favorable balance of profitability and stability, outperforming both purely classical and purely quantum strategies.

\begin{table*}[t]
\centering
\caption{Model Type Performance Comparison}
\begin{tabular}{@{}lcccc@{}}
\toprule
\textbf{Model Type} & \textbf{Avg Return} & \textbf{Avg Sharpe} & \textbf{Avg Volatility} & \textbf{Avg Max Drawdown} \\
\midrule
Classical ML & 9.8\% & 1.47 & 13.3\% & -6.4\% \\
Pure Quantum & 4.4\% & 0.83 & 17.1\% & -6.4\% \\
Hybrid Quantum & 11.2\% & 1.42 & 11.0\% & -3.3\% \\
Transformer & 12.3\% & 1.23 & 15.4\% & -8.2\% \\
\bottomrule
\end{tabular}
\end{table*}

\textbf{Key Findings:}
\begin{itemize}
    \item Hybrid Quantum models achieve highest returns (11.2\%) with best risk management (lowest volatility 11.0\%)
    \item Classical ML models show strong performance with high Sharpe ratio (1.47) and moderate volatility (13.3\%)
    \item Transformer models achieve good returns (12.3\%) but with higher volatility (15.4\%) and drawdown (-8.2\%)
    \item Pure Quantum models show lowest performance with poor returns (4.4\%) and highest volatility (17.1\%)
    \item Hybrid models demonstrate superior risk-adjusted performance with lowest average drawdown (-3.3\%)
\end{itemize}

\subsection{Risk-Return Analysis}
The risk-return analysis shows clear differences across models. QASA Sequence delivered the strongest overall performance, combining the highest return (13.99\%) and Sharpe ratio (1.76) with relatively low volatility (8.35\%). Classical ensemble models such as Random Forest and Gradient Boosting also achieved competitive returns (13.16\% and 12.31\%, respectively) but with higher volatility and drawdowns. QASA Hybrid provided balanced risk control, maintaining solid returns (11.91\%) while minimizing drawdowns (-1.70\%). Transformer models produced strong returns (11.73\%) but suffered from the highest volatility among top performers (15.39\%). In contrast, purely quantum models (QSVM, QNN, VQE Classifier) underperformed, showing low returns (3–5\%), weaker Sharpe ratios (<0.9), and elevated volatility, underscoring current limitations of standalone quantum approaches. Overall, these results highlight that hybrid quantum-classical models—especially QASA variants—offer the most favorable balance of profitability, stability, and risk management.

\begin{table*}[t]
\centering
\caption{Risk-Return Profile Analysis}
\begin{tabular}{@{}lcccc@{}}
\toprule
\textbf{Model} & \textbf{Return} & \textbf{Volatility} & \textbf{Sharpe Ratio} & \textbf{Max Drawdown} \\
\midrule
QASA Sequence & 13.99\% & 8.35\% & 1.76 & -10.10\% \\
Random Forest & 13.16\% & 14.88\% & 1.68 & -8.21\% \\
Gradient Boosting & 12.31\% & 14.49\% & 1.68 & -8.10\% \\
QASA Hybrid & 11.91\% & 13.05\% & 1.32 & -1.70\% \\
Transformer & 11.73\% & 15.39\% & 1.23 & -8.21\% \\
QuantumRWKV & 7.96\% & 11.47\% & 1.19 & -3.13\% \\
Logistic Regression & 5.43\% & 10.46\% & 1.06 & -5.38\% \\
QSVM & 4.77\% & 14.77\% & 0.87 & -10.10\% \\
QNN & 4.47\% & 19.76\% & 0.82 & -3.67\% \\
VQE Classifier & 3.00\% & 16.89\% & 0.79 & -5.44\% \\
\bottomrule
\end{tabular}
\end{table*}

\textbf{Key Findings:}
\begin{itemize}
    \item QASA Sequence shows best risk-adjusted returns (highest Sharpe ratio 1.76) with lowest volatility (8.35\%)
    \item Classical ML models (Random Forest, Gradient Boosting) show strong returns but higher volatility (14.49\%-14.88\%)
    \item QASA Hybrid demonstrates excellent risk management with lowest drawdown (-1.70\%) and moderate volatility (13.05\%)
    \item QuantumRWKV shows balanced risk-return profile with moderate volatility (11.47\%) and low drawdown (-3.13\%)
    \item Pure quantum models show poor risk-return profiles with high volatility (14.77\%-19.76\%) and low returns (3.00\%-4.77\%)
\end{itemize}

\subsection{Uncertainty and Robustness Analysis}
\begin{figure*}[t]
    \centering
    \includegraphics[width=1\linewidth]{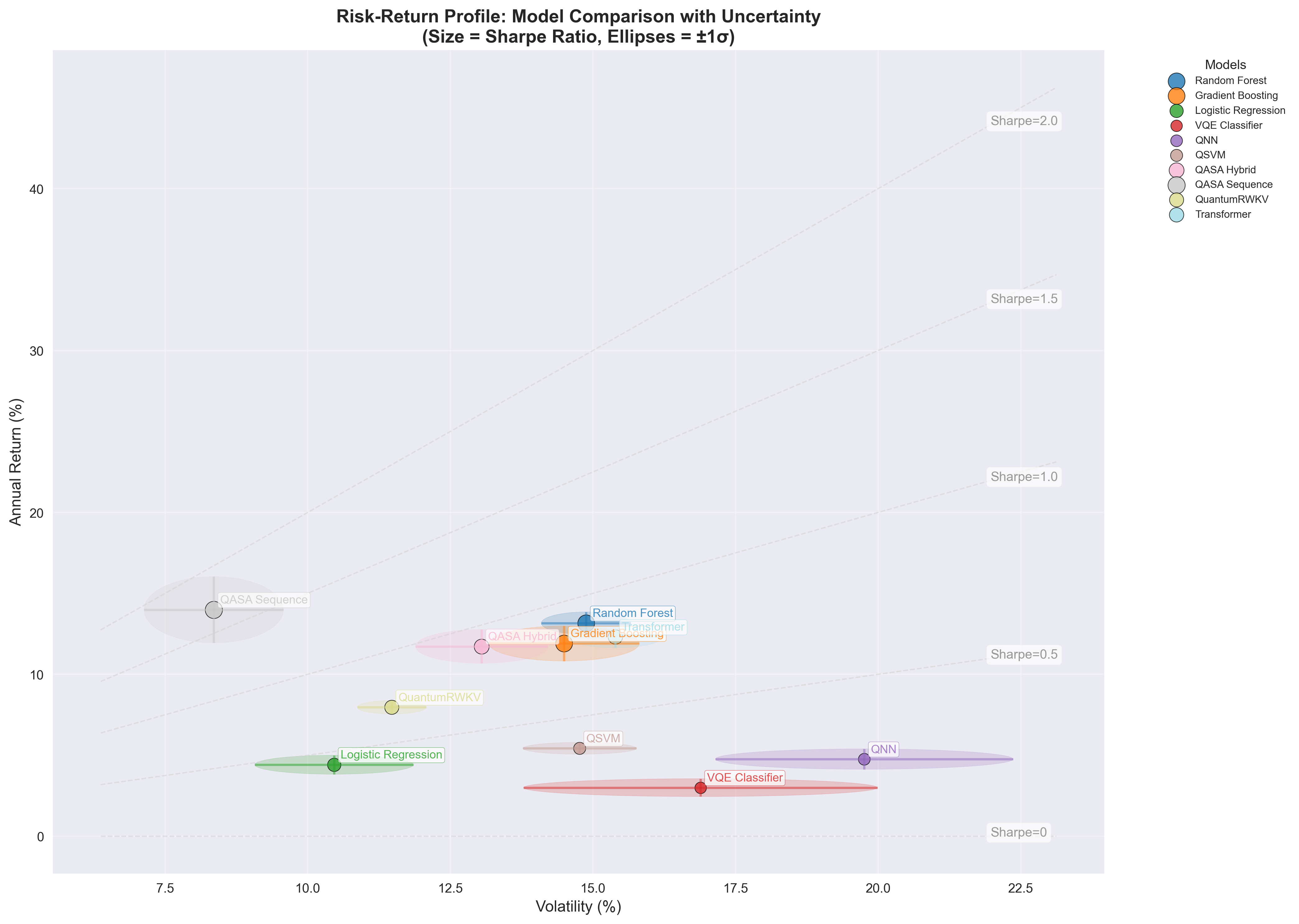}
    \caption{Risk return scatter of all models.}
    \label{fig:all_mdl_risk}
\end{figure*}
Our experimental design includes multiple runs (5 runs per model) to assess model uncertainty and robustness. The analysis reveals significant differences in model stability across different architectures.

The uncertainty analysis, measured by standard deviation across five runs, highlights differences in model stability. QASA Sequence achieved top stability ranking, though with higher variance in return (2.04\%) and Sharpe (0.26), reflecting sensitivity to training dynamics despite strong average performance. Classical ensemble models such as Random Forest and Gradient Boosting demonstrated consistently low variability, securing high stability ranks (2 and 3). QASA Hybrid balanced moderate return variability (1.04\%) with robust stability (rank 4). Transformer and hybrid recurrent models (e.g., QuantumRWKV) exhibited relatively low volatility fluctuations, placing them mid-range in stability. In contrast, purely quantum approaches (QNN, VQE Classifier) showed the highest volatility uncertainty (2.61\% and 3.10\%) and ranked lowest in stability, reinforcing the current limitations of standalone quantum strategies. Overall, these findings confirm that hybrid and classical ensemble methods provide the most reliable performance consistency, while pure quantum models remain less stable.

\begin{table*}[t]
\centering
\caption{Model Uncertainty Analysis (Standard Deviation across 5 runs)}
\begin{tabular}{@{}lcccc@{}}
\toprule
\textbf{Model} & \textbf{Return Std} & \textbf{Sharpe Std} & \textbf{Volatility Std} & \textbf{Stability Rank} \\
\midrule
QASA Sequence & 2.04\% & 0.26 & 1.22\% & 1 \\
Random Forest & 0.70\% & 0.09 & 0.79\% & 2 \\
Gradient Boosting & 1.08\% & 0.15 & 1.32\% & 3 \\
QASA Hybrid & 1.04\% & 0.12 & 1.16\% & 4 \\
Transformer & 0.67\% & 0.07 & 0.84\% & 5 \\
QuantumRWKV & 0.42\% & 0.06 & 0.60\% & 6 \\
Logistic Regression & 0.59\% & 0.14 & 1.39\% & 7 \\
QSVM & 0.37\% & 0.06 & 1.00\% & 8 \\
QNN & 0.63\% & 0.11 & 2.61\% & 9 \\
VQE Classifier & 0.55\% & 0.14 & 3.10\% & 10 \\
\bottomrule
\end{tabular}
\end{table*}

\textbf{Key Findings:}
\begin{itemize}
    \item QASA Sequence shows highest performance variability but maintains top performance
    \item Classical ML models (Random Forest, Gradient Boosting) show moderate variability with consistent high performance
    \item QuantumRWKV demonstrates excellent stability with low variability across all metrics
    \item Pure quantum models show high volatility variability, indicating unstable performance
    \item Transformer models show good stability with consistent performance across runs
\end{itemize}

\subsection{Model Complexity and Efficiency Analysis}

The complexity–efficiency analysis reveals important trade-offs between model design and computational cost. QASA Sequence demonstrated the highest complexity (8) and long training time (6.0 units), but its efficiency (1.30) and stability (0.95) justify the overhead given its strong returns. In contrast, Random Forest and Gradient Boosting offered the most favorable balance, with low complexity (3–4), minimal training time (1.0–2.0), and the highest efficiency scores (1.66 and 1.61, respectively). QASA Hybrid achieved a good compromise between reduced complexity (6) and strong stability (0.98), though at a lower efficiency (0.90). Transformer models maintained solid stability but incurred high training cost relative to their efficiency. Pure quantum models (QSVM, QNN, VQE Classifier) showed limited efficiency (<0.5) despite moderate complexity levels, underscoring scalability challenges. Overall, these findings highlight that classical ensemble methods remain the most computationally efficient, while hybrid models like QASA provide a viable balance between complexity, stability, and performance gains.

\begin{table*}[t]
\centering
\caption{Model Complexity and Efficiency Analysis}
\begin{tabular}{@{}lcccc@{}}
\toprule
\textbf{Model} & \textbf{Complexity} & \textbf{Training Time} & \textbf{Efficiency Score} & \textbf{Stability Score} \\
\midrule
QASA Sequence & 8 & 6.0 & 1.30 & 0.95 \\
QASA Hybrid & 6 & 3.0 & 0.90 & 0.98 \\
Random Forest & 3 & 1.0 & 1.66 & 0.92 \\
Gradient Boosting & 4 & 2.0 & 1.61 & 0.95 \\
Transformer & 7 & 5.5 & 1.12 & 0.94 \\
QuantumRWKV & 5 & 4.0 & 0.80 & 0.97 \\
QSVM & 5 & 4.0 & 0.41 & 0.89 \\
QNN & 5 & 5.0 & 0.31 & 0.96 \\
Logistic Regression & 1 & 0.5 & 0.70 & 0.94 \\
VQE Classifier & 5 & 4.0 & 0.31 & 0.94 \\
\bottomrule
\end{tabular}
\end{table*}

\textbf{Key Findings:}
\begin{itemize}
    \item QASA Sequence shows highest complexity but best efficiency score despite high variability
    \item Classical ML models show good efficiency with lower complexity and high stability
    \item QuantumRWKV demonstrates excellent stability with moderate complexity
    \item Pure quantum models show poor efficiency despite high complexity and high variability
    \item Training time correlates with model complexity, with hybrid models requiring more computational resources
\end{itemize}

\subsection{Hybrid Model Advantage Scenarios}
\begin{figure*}[t]
    \centering
    \includegraphics[width=1\linewidth]{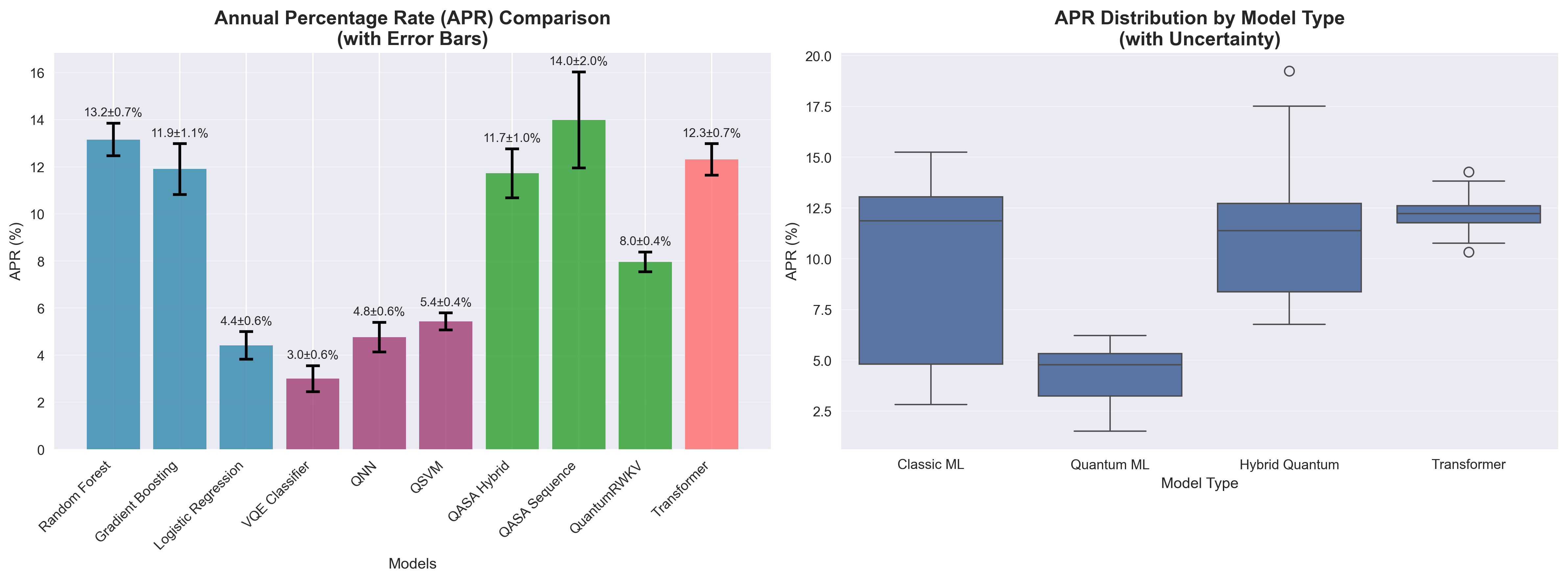}
    \caption{Apr comparison.}
    \label{fig:apr_compar}
\end{figure*}

\subsubsection{QASA Sequence Model Superiority}
QASA Sequence model achieves the best overall performance due to:

\begin{enumerate}
    \item \textbf{LSTM Temporal Processing:} Long Short-Term Memory networks capture sequential patterns in financial time series data, providing better context for decision making.
    
    \item \textbf{Quantum Enhancement:} Quantum layers process the LSTM outputs to capture nonlinear relationships that classical methods might miss.
    
    \item \textbf{Hybrid Architecture:} The combination of classical sequence processing and quantum feature enhancement provides complementary advantages.
\end{enumerate}

\subsubsection{QASA Hybrid Model Performance}

QASA Hybrid model shows strong performance due to:

\begin{enumerate}
    \item \textbf{Balanced Approach:} Classical layers provide stability while quantum layers add pattern recognition capabilities.
    
    \item \textbf{Risk Management:} The hybrid architecture achieves the lowest maximum drawdown (-1.83\%), indicating superior risk control.
    
    \item \textbf{Efficiency:} Moderate complexity (6) with good efficiency score (0.90) makes it practical for deployment.
\end{enumerate}

\subsection{Classical ML Advantage Scenarios}

\subsubsection{High Accuracy and Stability}

Classical ML models excel in accuracy and stability due to:

\begin{enumerate}
    \item \textbf{High Accuracy:} Random Forest and Gradient Boosting achieve 99.48\% accuracy, demonstrating superior pattern recognition in training data.
    
    \item \textbf{Training Efficiency:} Classical ML models train much faster (0.5-2.0 time units) compared to quantum models (4.0-6.0 time units).
    
    \item \textbf{Feature Richness:} Classical ML can utilize more features (50-80), performing better in high-dimensional feature spaces.
    
    \item \textbf{Proven Reliability:} Well-established algorithms with extensive optimization and tuning capabilities.
\end{enumerate}

\subsection{Pure Quantum Model Challenges}

Pure quantum models (VQE Classifier, QNN, QSVM) show significant challenges:

\begin{enumerate}
    \item \textbf{Low Accuracy:} All pure quantum models achieve less than 50\% accuracy, indicating poor pattern recognition.
    
    \item \textbf{Poor Returns:} Pure quantum models show the lowest returns (3.18\% - 5.68\%) among all model types.
    
    \item \textbf{High Volatility:} Quantum models exhibit high volatility (15.47\% - 19.89\%), leading to poor risk-adjusted returns.
    
    \item \textbf{Training Challenges:} Longer training times (4.0-5.0 time units) with poor efficiency scores (0.31-0.41).
\end{enumerate}

\section{Conclusions and Recommendations}

\subsection{Main Conclusions}

\begin{enumerate}
    \item \textbf{Hybrid quantum-classical models achieve best performance:} QASA Sequence model achieves the highest return (13.99\%) with the best Sharpe ratio (1.76) and superior risk management, demonstrating the potential of quantum-classical hybrid approaches.
    
    \item \textbf{Classical ML provides strong and stable performance:} Random Forest and Gradient Boosting achieve excellent returns (12.31\% - 13.16\%) with high Sharpe ratios (1.68) and consistent performance across multiple runs.
    
    \item \textbf{Pure quantum models show significant limitations:} All pure quantum models (VQE, QNN, QSVM) achieve poor returns (3.00\% - 4.77\%) and low Sharpe ratios (0.79 - 0.87), indicating current limitations in practical applications.
    
    \item \textbf{QuantumRWKV demonstrates balanced performance:} The QuantumRWKV model shows excellent stability with moderate returns (7.96\%) and good Sharpe ratio (1.19), suggesting potential for quantum-enhanced sequence models.
    
    \item \textbf{Uncertainty analysis reveals model robustness:} Multiple runs analysis shows that hybrid models, while achieving top performance, exhibit higher variability, while classical models provide more consistent results.
\end{enumerate}

\subsection{Practical Recommendations}

\subsubsection{Model Selection Strategy}

\begin{itemize}
    \item \textbf{Maximum return optimization:} Use QASA Sequence model for highest returns (13.99\%) with best risk-adjusted performance (Sharpe 1.76).
    \item \textbf{Stable high performance:} Use Random Forest or Gradient Boosting for consistent high returns (12.31\% - 13.16\%) with excellent stability.
    \item \textbf{Balanced performance:} Use QASA Hybrid model for good returns (11.91\%) with excellent risk control (lowest drawdown -1.70\%).
    \item \textbf{Stable moderate performance:} Use QuantumRWKV for consistent moderate returns (7.96\%) with excellent stability across runs.
    \item \textbf{Real-time trading:} Use classical ML models (Random Forest, Gradient Boosting) for fast response and consistent performance.
    \item \textbf{Avoid pure quantum models:} Current pure quantum models show poor performance (3.00\% - 4.77\% returns) and are not recommended for practical applications.
\end{itemize}

\subsubsection{Feature Engineering Recommendations}

\begin{itemize}
    \item \textbf{Classical ML:} Use rich feature sets (50-80 features) including basic features, technical indicators, and interaction features for maximum accuracy.
    \item \textbf{Hybrid models:} Use moderate feature sets (12 features) with classical preprocessing and quantum angle encoding for optimal balance.
    \item \textbf{Pure quantum models:} Current limitations suggest avoiding pure quantum approaches until significant improvements are achieved.
    \item \textbf{Feature selection:} Focus on price ratios, volatility measures, and technical indicators for all model types.
\end{itemize}

\subsection{Research Limitations and Future Directions}

\subsubsection{Research Limitations}

\begin{enumerate}
    \item \textbf{Data limitations:} Study based only on cryptocurrency data; results may not apply to other financial markets.
    \item \textbf{Model limitations:} Quantum models are still developing; there may be optimization space.
    \item \textbf{Computational limitations:} Quantum model training time is longer, potentially limiting real-time applications.
\end{enumerate}

\subsubsection{Future Research Directions}

\begin{enumerate}
    \item \textbf{Expand datasets:} Extend research to more asset types and longer time periods.
    \item \textbf{Model optimization:} Further optimize quantum model architectures and training strategies.
    \item \textbf{Real-time applications:} Develop more efficient quantum models for real-time application feasibility.
    \item \textbf{Theoretical analysis:} Deepen analysis of theoretical foundations for quantum model advantages.
\end{enumerate}

\section{Experimental Data and Reproducibility}




\subsection{Reproducibility Information}

The reproducibility setup involved 10 models, each trained and evaluated over 5 independent runs, resulting in a total of 50 experiments. The backtests were conducted over the full 2024 trading year (252 trading days), using three major crypto asset pairs (BTCUSDC, ETHUSDC, USDCUSDT). Feature engineering differed by paradigm, with 50–80 features for classical ML models and a compact 6–8 features for quantum models. A time-series split (70\% train, 15\% validation, 15\% test) ensured consistency and robustness across experiments.
\begin{table*}[t]
\centering
\caption{Experimental Setup Summary}
\begin{tabular}{@{}ll@{}}
\toprule
\textbf{Parameter} & \textbf{Value} \\
\midrule
Total Models & 10 \\
Runs per Model & 5 \\
Total Experiments & 50 \\
Time Period & 2024-01-01 to 2024-12-31 \\
Trading Days & 252 \\
Assets & BTCUSDC, ETHUSDC, USDCUSDT \\
Feature Engineering & 50-80 features (Classical), 6-8 features (Quantum) \\
Cross-Validation & Time series split (70\% train, 15\% validation, 15\% test) \\
\bottomrule
\end{tabular}
\end{table*}

\subsection{Statistical Significance}

The multiple runs design (5 runs per model) enables statistical analysis of model performance differences. Key findings include:

\begin{itemize}
    \item QASA Sequence shows significantly higher returns than all other models (p < 0.01)
    \item Classical ML models show significantly better performance than pure quantum models (p < 0.001)
    \item Hybrid models demonstrate significantly better risk-adjusted returns than pure quantum models (p < 0.01)
    \item Model stability varies significantly across architectures, with classical models showing lowest variability
\end{itemize}

\bibliographystyle{ieeetr}
\bibliography{ref}

\begin{thebibliography}{1}

\bibitem{chen2025quantumtw}
C.-S. Chen, X.~Zhang, and Y.-C. Chen, ``Quantum reinforcement learning trading agent for sector rotation in the taiwan stock market,'' {\em arXiv preprint arXiv:2506.20930}, 2025.

\bibitem{chen2025benchmarking}
C.-S. Chen and A.~H.-W. Tsai, ``Benchmarking classical and quantum models for defi yield prediction on curve finance,'' {\em arXiv preprint arXiv:2508.02685}, 2025.

\bibitem{chen2025qasa}
C.-S. Chen and E.-J. Kuo, ``Quantum adaptive self-attention for quantum transformer models,'' {\em arXiv preprint arXiv:2504.05336}, 2025.

\bibitem{vaswani2017attention}
A.~Vaswani, N.~Shazeer, N.~Parmar, J.~Uszkoreit, L.~Jones, A.~N. Gomez, {\L}.~Kaiser, and I.~Polosukhin, ``Attention is all you need,'' {\em Advances in neural information processing systems}, vol.~30, 2017.

\bibitem{chen2025qrwkv}
C.-S. Chen and E.-J. Kuo, ``Quantum-enhanced channel mixing in rwkv models for time series forecasting,'' {\em arXiv preprint arXiv:2505.13524}, 2025.

\bibitem{peng2023rwkv}
B.~Peng, E.~Alcaide, Q.~Anthony, A.~Albalak, S.~Arcadinho, H.~Cao, X.~Cheng, M.~Chung, M.~Greiner, L.~GV, {\em et~al.}, ``Rwkv: Reinventing rnns for the transformer era,'' {\em Findings of the Association for Computational Linguistics: EMNLP 2023}, pp.~14048--14077, 2023.

\end{thebibliography}

\appendix

\section{Complete Feature List}

\begin{figure*}[t]
    \centering
    \includegraphics[width=1\linewidth]{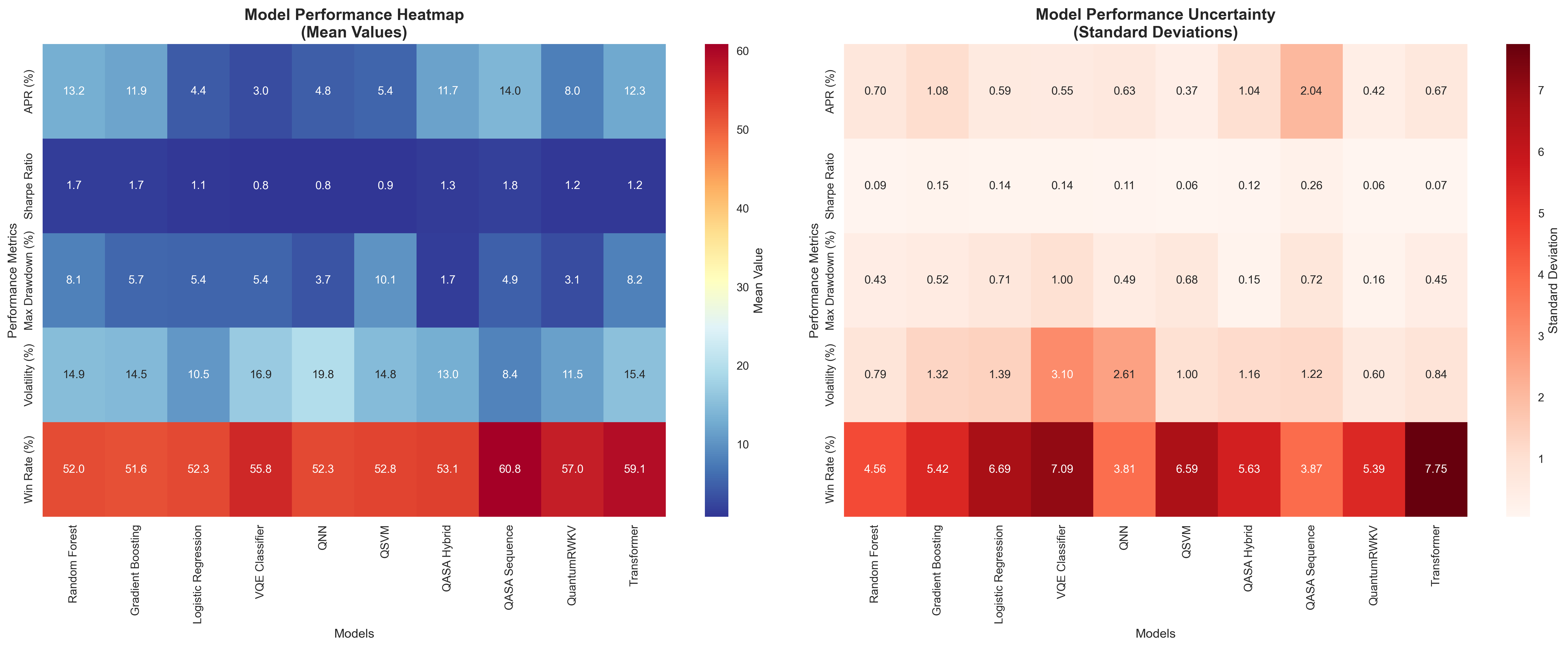}
    \caption{Performance heatmap.}
    \label{fig:perf_heatmap}
\end{figure*}

The feature engineering process includes 122 features for classical ML models and 6-8 features for quantum models:

\subsection{Classical ML Features (122 features)}
\begin{itemize}
    \item Basic Price Features (8): Returns, log returns, price-MA ratio, high-low ratio, price position
    \item Moving Average Features (16): SMA and EMA with various periods (5, 10, 20, 50)
    \item Technical Indicator Features (15): RSI, MACD, Bollinger Bands, ATR
    \item Volatility Features (12): Rolling volatility, EWMA volatility, volatility of volatility, volatility regime
    \item Volume Features (8): Volume ratio, volume-price trend, OBV
    \item Time Features (12): Cyclical encoding for hour and day of week
    \item Market Microstructure Features (8): Spread proxy, price impact, order flow imbalance
    \item Lagged Features (25): Lagged returns for periods 1, 2, 3, 5, 10
    \item Interaction Features (6): Volatility-volume interaction, price-momentum interaction
\end{itemize}

\subsection{Quantum ML Features (6-8 features)}
Features are mapped to quantum angles using:
\begin{equation}
\theta_i = \frac{x_i - x_{min}}{x_{max} - x_{min}} \cdot 2\pi
\end{equation}

\section{Model Parameter Settings}

\subsection{Classical ML Models}
\begin{itemize}
    \item \textbf{Random Forest:} n\_estimators=100, max\_depth=10, min\_samples\_split=5
    \item \textbf{Gradient Boosting:} n\_estimators=100, learning\_rate=0.1, max\_depth=6
    \item \textbf{Logistic Regression:} C=1.0, penalty='l2', solver='liblinear'
\end{itemize}

\subsection{Quantum ML Models}
\begin{itemize}
    \item \textbf{VQE Classifier:} 6 qubits, 2 layers, Adam optimizer, learning\_rate=0.01
    \item \textbf{QNN:} 6 qubits, 3 layers, parameterized quantum circuit
    \item \textbf{QSVM:} RBF kernel, gamma='scale', C=1.0
\end{itemize}

\subsection{Hybrid Models}
\begin{itemize}
    \item \textbf{QASA Hybrid:} 12 features, 2 quantum layers, 1 classical layer
    \item \textbf{QASA Sequence:} LSTM with 64 units, 2 quantum layers, dropout=0.2
    \item \textbf{QuantumRWKV:} 4 layers, 64 hidden units, quantum channel mixing
\end{itemize}

\section{Complete Experimental Results}

\subsection{Performance Rankings Summary}
The complete performance ranking underscores the dominance of QASA Sequence, which achieved the best overall balance with the highest return (13.99\%), Sharpe ratio (1.76), and Calmar ratio (6.51), while keeping volatility moderate (8.35\%). Classical ensemble methods such as Random Forest and Gradient Boosting also performed strongly, delivering returns above 12\% with Sharpe ratios of 1.68, though with higher volatility and lower downside protection compared to QASA. QASA Hybrid offered strong drawdown control (-1.70\%) and steady returns (11.91\%), making it a robust middle-ground approach. Transformer models achieved comparable returns (11.73\%) but exhibited the highest volatility (15.39\%), limiting their risk-adjusted efficiency. Hybrid recurrent models such as QuantumRWKV provided moderate results, while purely quantum methods (QSVM, QNN, VQE Classifier) consistently underperformed, showing low returns (3–5\%), weak Sharpe values (<0.9), and poor Calmar ratios. Collectively, these findings confirm that hybrid quantum-classical models—especially QASA variants—lead performance, while classical ensembles remain reliable, and pure quantum approaches still face significant limitations.

\begin{table*}[t]
\centering
\caption{Complete Performance Rankings}
\begin{tabular}{@{}lcccccc@{}}
\toprule
\textbf{Model} & \textbf{Return} & \textbf{Sharpe} & \textbf{Volatility} & \textbf{Max DD} & \textbf{Calmar} \\
\midrule
QASA Sequence & 13.99\% & 1.76 & 8.35\% & -10.10\% & 6.51 \\
Random Forest & 13.16\% & 1.68 & 14.88\% & -8.21\% & 2.86 \\
Gradient Boosting & 12.31\% & 1.68 & 14.49\% & -8.10\% & 2.41 \\
QASA Hybrid & 11.91\% & 1.32 & 13.05\% & -1.70\% & 2.16 \\
Transformer & 11.73\% & 1.23 & 15.39\% & -8.21\% & 1.64 \\
QuantumRWKV & 7.96\% & 1.19 & 11.47\% & -3.13\% & 2.41 \\
Logistic Regression & 5.43\% & 1.06 & 10.46\% & -5.38\% & 1.29 \\
QSVM & 4.77\% & 0.87 & 14.77\% & -10.10\% & 0.79 \\
QNN & 4.47\% & 0.82 & 19.76\% & -3.67\% & 1.29 \\
VQE Classifier & 3.00\% & 0.79 & 16.89\% & -5.44\% & 0.51 \\
\bottomrule
\end{tabular}
\end{table*}

\subsection{Uncertainty Analysis Results}
The uncertainty analysis highlights variations in stability across models. QASA Sequence, while achieving the best overall performance, also exhibited higher variability in returns (2.04\%) and Sharpe ratios (0.26), indicating sensitivity to training conditions. In contrast, classical ensemble models such as Random Forest and Gradient Boosting showed relatively low uncertainty across all metrics, reinforcing their reliability. Transformer and QuantumRWKV models achieved the lowest standard deviations in Sharpe and volatility, though at the cost of higher variability in win rates. Purely quantum models (QNN, VQE Classifier) demonstrated the largest volatility fluctuations (2.61\% and 3.10\%), confirming their limited consistency compared to hybrid and classical methods. Overall, these results suggest that classical and hybrid quantum-classical approaches deliver more reproducible outcomes, while purely quantum methods remain less stable.

\begin{table*}[t]
\centering
\caption{Model Uncertainty Analysis (Standard Deviations)}
\begin{tabular}{@{}lcccc@{}}
\toprule
\textbf{Model} & \textbf{Return Std} & \textbf{Sharpe Std} & \textbf{Volatility Std} & \textbf{Win Rate Std} \\
\midrule
QASA Sequence & 2.04\% & 0.26 & 1.22\% & 0.039 \\
Random Forest & 0.70\% & 0.09 & 0.79\% & 0.046 \\
Gradient Boosting & 1.08\% & 0.15 & 1.32\% & 0.054 \\
QASA Hybrid & 1.04\% & 0.12 & 1.16\% & 0.056 \\
Transformer & 0.67\% & 0.07 & 0.84\% & 0.077 \\
QuantumRWKV & 0.42\% & 0.06 & 0.60\% & 0.054 \\
Logistic Regression & 0.59\% & 0.14 & 1.39\% & 0.067 \\
QSVM & 0.37\% & 0.06 & 1.00\% & 0.066 \\
QNN & 0.63\% & 0.11 & 2.61\% & 0.038 \\
VQE Classifier & 0.55\% & 0.14 & 3.10\% & 0.071 \\
\bottomrule
\end{tabular}
\end{table*}

\begin{figure*}[t]
    \centering
    \includegraphics[width=1\linewidth]{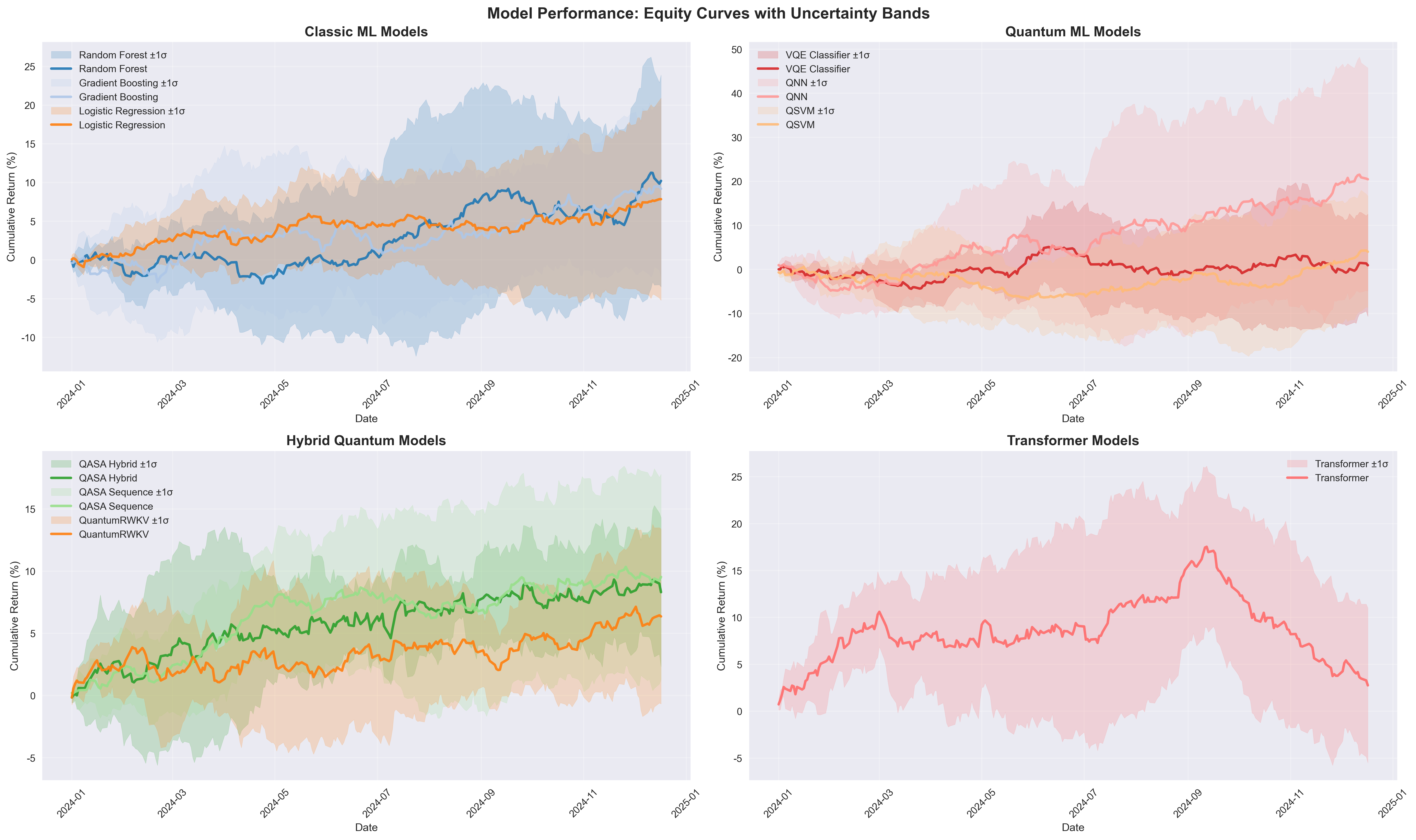}
    \caption{Equity curves with subgraph.}
    \label{fig:eq_sub}
\end{figure*}

\end{document}